\documentclass{article}
\usepackage{spconf,amsmath,graphicx}

\usepackage{enumitem}
\setlist{nosep, leftmargin=14pt}

\usepackage[nolist]{acronym}
\usepackage{dirtytalk}
\usepackage{hyperref}
\usepackage[capitalise, noabbrev]{cleveref}
\usepackage{adjustbox}
\usepackage{subcaption}
\usepackage{amssymb}
\usepackage{svg}
\usepackage{amsmath}

\usepackage{tikz}
\usetikzlibrary{shapes.geometric, positioning, calc, arrows, shapes.misc, decorations.pathreplacing, calligraphy}
\usepackage{rotating} 
\usepackage[]{xcolor}
\usepackage{scalerel}

\usepackage{siunitx}
\DeclareSIUnit\pixel{px}

\usepackage{xspace}
\makeatletter
\DeclareRobustCommand\onedot{\futurelet\@let@token\@onedot}
\def\@onedot{\ifx\@let@token.\else.\null\fi\xspace}
\def\etal{\emph{et al}\onedot}

\newcommand{\barlow}[0]{\textit{Barlow Twin}}
\newcommand{\twins}[0]{\textit{Barlow Twins}}
\newcommand{\triplets}[0]{\textit{Barlow Triplets}}

\title{Mind the Gap: Scanner-induced domain shifts pose challenges for representation learning in histopathology}

\name{
    \begin{tabular}{c}
        Frauke Wilm$^{1,2}$ ,  Marco Fragoso$^{3}$, Christof A. Bertram$^{4}$, Nikolas Stathonikos$^{5}$, Mathias Öttl$^{1}$, \\ Jingna Qiu$^{2}$, Robert Klopfleisch$^{3}$, Andreas Maier$^{1}$, Marc Aubreville$^{6, \dagger}$, Katharina Breininger$^{2, \dagger}$\thanks{$^\dagger$ Shared senior authors.}
    \end{tabular}
}

\address{$^{1}$ Pattern Recognition Lab, Friedrich-Alexander-Universit\"at Erlangen-N\"urnberg, Germany\\
$^{2}$ Department AIBE, Friedrich-Alexander-Universit\"at Erlangen-N\"urnberg, Germany\\
$^{3}$ Institute of Veterinary Pathology, Freie Universit\"at Berlin, Germany\\
$^{4}$ Institute of Pathology, University of Veterinary Medicine, Vienna, Austria\\
$^{5}$ Pathology Department, University Medical Centre Utrecht, The Netherlands\\
$^{6}$ Technische Hochschule Ingolstadt, Ingolstadt, Germany} 

\begin{document}
%
\maketitle
\begin{abstract}
Computer-aided systems in histopathology are often challenged by various sources of domain shift that impact the performance of these algorithms considerably. We investigated the potential of using self-supervised pre-training to overcome scanner-induced domain shifts for the downstream task of tumor segmentation. For this, we present the \triplets{} to learn scanner-invariant representations from a multi-scanner dataset with local image correspondences. We show that self-supervised pre-training successfully aligned different scanner representations, which, interestingly only results in a limited benefit for our downstream task. We thereby provide insights into the influence of scanner characteristics for downstream applications and contribute to a better understanding of why established self-supervised methods have not yet shown the same success on histopathology data as they have for natural images.
\end{abstract}
\section{Introduction}
\label{sec:intro}
Machine learning-based image analysis has experienced an upswing in histopathology, to which the ease of creating digital \acp{wsi} using slide scanners greatly contributed. Whilst computer-aided systems often excel within their training domain - sometimes even surpassing the performance of experienced pathologists~\cite{aubreville2020} - their performance considerately decreases for out-of-distribution samples~\cite{aubreville2021, lafarge2019, stacke2020}. Representation learning has previously shown great success for domain generalization~\cite{motiian2017} by pre-training models with cross-domain image pairs and aligning their representations directly in the embedding space. These cross-domain image pairs are usually created without supervision through augmentations that simulate possible domain shifts. Recent work has shown that self-supervised pre-training can help to extract task-relevant information and increase the performance for downstream tasks even without supervision for these tasks at the pre-training stage~\cite{tsai2020, tian2020,lee2021}. These works, however, see a high dependency of the downstream performance on the amount and type of information the constructed views share. If two views share too little task-relevant information, performance on the downstream task can degrade, whilst if they share too many tasks-irrelevant (nuisance) features, self-supervised pre-training can lead to the extraction of shortcut features~\cite{tian2020, stacke2021}. Based on these observations, Tian \etal stated the \textit{InfoMin} objective of finding a sweet spot where two views retain enough task-relevant information without sharing irrelevant nuisances \cite{tian2020}.

In this work, we try to find this sweet spot for creating views that help to mitigate scanner-induced domain shifts. For this, we utilized a multi-scanner dataset with local image correspondences. By digitizing the same sample with multiple slide scanning systems, we preserved task-relevant features (e.g. tissue morphology) while introducing \say{intrinsic} domain shifts that might exceed apparent differences such as color or resolution. For representation learning, we adapted the \barlow{}~\cite{zbontar2021} architecture to accommodate multi-scanner tuples and evaluated this use of self-supervised pre-training for the downstream task of tumor segmentation on canine skin cancer specimens. Our experiments thereby extend the work of Stacke \etal who have applied representation learning for the task of tissue phenotyping in the presence of cross-organ and cross-laboratory domain shifts. The authors, however, faced challenges when applying established methods of self-supervised learning to histopathology and argued that \say{the subtleties of the difference between classes in histology data makes it more challenging to find effective augmentations}~\cite{stacke2021}, which we circumvent by using a multi-scanner dataset that inherently possesses these differences. 

Even though we show that self-supervised pre-training helped to align the scanner embeddings, we only observed a limited benefit for our downstream task. We suspect that feature alignment enhanced some scanner characteristics that might be beneficial for tumor segmentation, but that the small inter-class variance of histopathology limits the benefit for the downstream task. We thereby contribute to a better understanding of why self-supervised learning has not yet shown the same potential for histopathology data as it has for natural images~\cite{stacke2021} and provide recommendations for using representation learning to approach the domain shift inherent in multi-scanner histopathology datasets.  

\section{Materials}
\begin{figure}[htb]
\begin{minipage}[b]{.2\linewidth}
  \centering
  \centerline{\includegraphics[width=1.7cm]{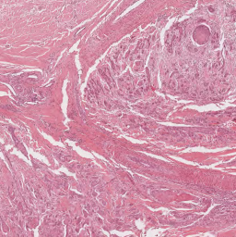}}
  \centerline{(a) CS2}\medskip
\end{minipage}%
\begin{minipage}[b]{.2\linewidth}
  \centering
  \centerline{\includegraphics[width=1.7cm]{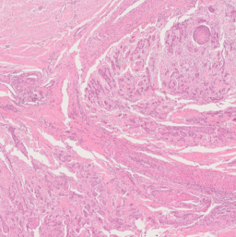}}
  \centerline{(b) NZ 210}\medskip
\end{minipage}%
\begin{minipage}[b]{.2\linewidth}
  \centering
  \centerline{\includegraphics[width=1.7cm]{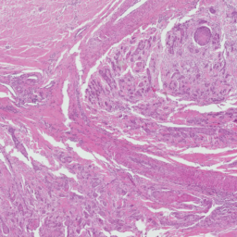}}
  \centerline{(c) NZ 2.0}\medskip
\end{minipage}%
\begin{minipage}[b]{.2\linewidth}
  \centering
  \centerline{\includegraphics[width=1.7cm]{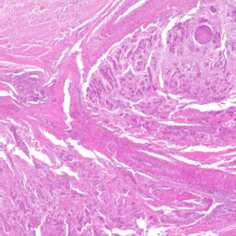}}
  \centerline{(d) P 1000}\medskip
\end{minipage}%
\begin{minipage}[b]{0.2\linewidth}
  \centering
  \centerline{\includegraphics[width=1.7cm]{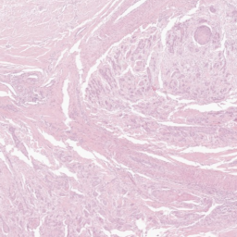}}
  \centerline{(e) GT 450}\medskip
\end{minipage}
\caption{Exemplary patch of the multi-scanner dataset.}
\label{fig:patches}
\end{figure}

Our experiments were performed on the \ac{scc} subset of the publicly available CATCH dataset~\cite{catch}. Use of these samples was approved by the local governmental authorities (State Office of Health and Social Affairs of Berlin, approval ID: StN 011/20). The specimens were originally digitized with the Aperio ScanScope CS2 (Leica, Germany) at a resolution of \SI{0.25}{\micro\meter\per\pixel} using a $40\times$ objective lens. To create a multi-scanner dataset with local correspondences, we digitized the samples with four additional slide scanners (exemplary patches in \cref{fig:patches}) : 
\begin{itemize}
    \item NanoZoomer S210 (Hamamatsu, Japan), \SI{0.22}{\micro\meter\per\pixel}
    \item NanoZoomer 2.0-HT (Hamamatsu, Japan), \SI{0.23}{\micro\meter\per\pixel}
    \item Pannoramic 1000 (3DHISTECH, Hungary), \SI{0.25}{\micro\meter\per\pixel}
    \item Aperio GT 450 (Leica, Germany), \SI{0.26}{\micro\meter\per\pixel} 
\end{itemize}%

All Aperio ScanScope CS2 \acp{wsi} were manually annotated for tumor and three skin tissue classes (epidermis, dermis, subcutis) using the open-source software SlideRunner~\cite{aubreville2018}. For automatic background detection, the images were converted to grayscale and non-annotated regions with a grayscale value above 235 were labeled as background. All remaining non-annotated regions were excluded from training and evaluation. Due to the image-to-image correspondences, all annotations could be transferred to the remaining scanners using the \ac{wsi} registration algorithm by Marzahl \etal~\cite{marzahl2021}. The registration success was visually validated by overlaying the transformed polygon annotations onto the \acp{wsi}. Due to severe scanning artifacts in at least one of the scans, six specimens were excluded from the dataset, resulting in a total of 220 \acp{wsi} (44 samples digitized with five scanners each). For algorithm development, a slide-level split into 30-5-9 train-validation-test was performed. The GT 450 and P 1000 scanners were selected as hold-out test scanners, to test the model's capability of generalizing to unseen domains. 

\section{Methods}
Given the multi-scanner dataset with local image correspondences, the goal was to extract scanner-invariant features for a subsequent tumor segmentation task. For this, we followed a two-step training pipeline: We first pre-trained a feature extractor in a self-supervised fashion and then switched to a fully-supervised training setup for the segmentation task.

For pre-training, we used the \barlow{} architecture~\cite{zbontar2021}, which originally uses a Siamese structure to create representations for two augmented versions of the same input image, that are then projected into a higher dimensional feature space $\mathbb{R}^d$, where the cross-correlation matrix $\mathcal{C}$ of both embeddings is computed. The \barlow{} loss $\mathcal{L_{BT}}$ enforces this matrix to be similar to the identity matrix:
\begin{equation}
\mathcal{L_{BT}} =  \sum_{i=0}^d (1-\mathcal{C_\mathit{ii}})^2 + \lambda \sum_{i=0}^d \sum_{\substack{j=0 \\ j \neq i}}^d \mathcal{C_\mathit{ij}}^2 \hspace{.5cm}
\end{equation}%
By enforcing high values on the main diagonal, features become invariant to the applied distortions (or in our case: the scanner-induced domain shift), whilst low values elsewhere disentangle features and thereby reduce redundancy.

We adapted the \twins{} to generalize to input tuples composed of one patch per training scanner. \Cref{fig:architecture} visualizes our adapted version for the case of three training scanners - the \textit{Barlow Triplets}. We performed a straightforward extension of the \barlow{} loss function to a \textit{Barlow Tuple} loss $\mathcal{L_{BT^\star}}$, that computes the joint loss of all unique pairs of a given input tuple as:
\begin{equation}
\mathcal{L_{BT^\star}} =  \frac{1}{K(n,2)} \sum_{k=1}^{K(n,2)} \mathcal{L_{BT,\ \mathit{k}}} \hspace{.5cm},
\end{equation}%
where $K(n,2) = \frac{n!}{(n-2)!2!}$ computes the number of unique pairs that can be constructed from $n$ scanner embeddings, e.g. $K(3,2) = \frac{6}{2}=3$ combinations for scanner triplets. 

\begin{figure}[htb]
\centering
\begin{minipage}[b]{1.0\linewidth}
  \centering
  \centerline{\input{figures/architecture_adpated.tex}}
\end{minipage}
\caption{\triplets{}. Figure adapted from~\cite{zbontar2021}.}
\label{fig:architecture}
\end{figure}

As encoder, we used a ResNet18~\cite{he2016} pre-trained on ImageNet~\cite{russakovsky2015}.  For the projector, we followed the implementation of Zbontar \etal and used three linear layers with a four-fold up-scaling in feature dimension~\cite{zbontar2021}, i.e. \num{2048} output units for a ResNet18 encoder. Similar to Zbontar \etal the first two linear layers were followed by batch normalization and rectified linear units.

After self-supervised pre-training, we used the annotation database to perform a fully-supervised segmentation training into background, tumor, and a third non-tumor class that combined all skin tissue classes. To construct a segmentation model, the encoder was connected to a decoding branch using skip connections in a U-Net-like fashion~\cite{ronneberger2015}. 

\newpage
\noindent \textbf{Training parameters.} The network was trained on $256\times256$ pixel-sized image patches. To cover more tissue context, we extracted the patches at a lower resolution of \SI{4}{\micro\meter\per\pixel}. For each epoch, we sampled 50 patches from each \ac{wsi} using an equal weighting of tumor and non-tumor and 10\% background patches. To increase the diversity of training data, this
guided selection of patches was repeated for each training epoch but kept unchanged for the validation set. All patches were z-score normalized using the mean and standard deviation of all tissue-containing areas of the CS2 training \acp{wsi}.

During self-supervised pre-training, we trained the encoder for 200 epochs, after which we observed model convergence. We used the Adam optimizer, a batch size of 64, and a cyclic learning rate with a maximum of \num{e-6}. Following Zbontar \etal~\cite{zbontar2021}, we selected $\lambda=\num{5e-3}$ for $\mathcal{L_{BT^\star}}$. After pre-training, we switched to a fully-supervised setup and trained the extended models for another 100 epochs using a cyclic learning rate with a maximum of \num{e-4}. For this, we used a reduced batch size of 8 (due to memory constraints) and a combination of cross-entropy and Dice loss~\cite{sudre2017}. Model selection was guided by the highest \ac{miou} score on the validation patches. 

\section{Experiments and Results}
To evaluate our representations, we compared the performance of our pre-trained encoder to a baseline initialized with ImageNet~\cite{russakovsky2015} weights for the downstream task of tumor segmentation. We evaluated two different settings: A single-domain approach, where only the CS2 patches were used for fully-supervised training of tumor segmentation (single), and a multi-domain approach, where the CS2 and both NanoZoomer scanners were used (multi). For each experiment, we repeated the fully-supervised training three times and averaged the test performance. For a fair comparison we used seeding to initialize the optimizers and data loaders to ensure that for each seed, all models obtained the same ordering of randomly sampled training patches and the same validation patches for the fully-supervised training stage. We used the same hyperparameters to train all models.   

\begin{figure}[htb]
\centering
\begin{subfigure}[c]{0.9\linewidth}
\subcaption{First stage: \textit{Barlow Triplet} pre-training}
\includegraphics[width=\linewidth]{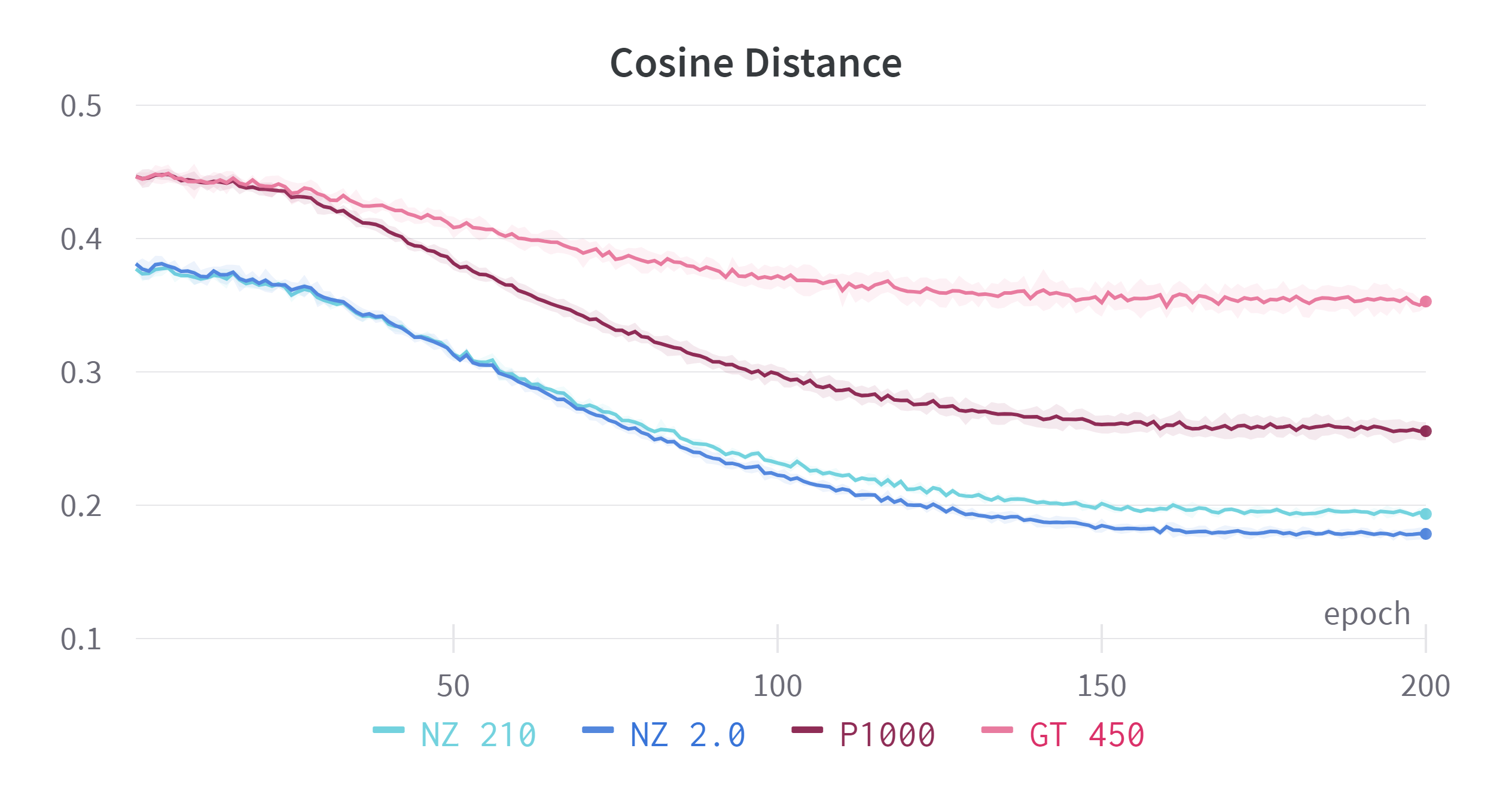}
\label{subfig:pretraining}
\end{subfigure}
\begin{subfigure}[c]{0.9\linewidth}
\subcaption{Second stage: Downstream task (scanner average)}
\includegraphics[width=\linewidth]{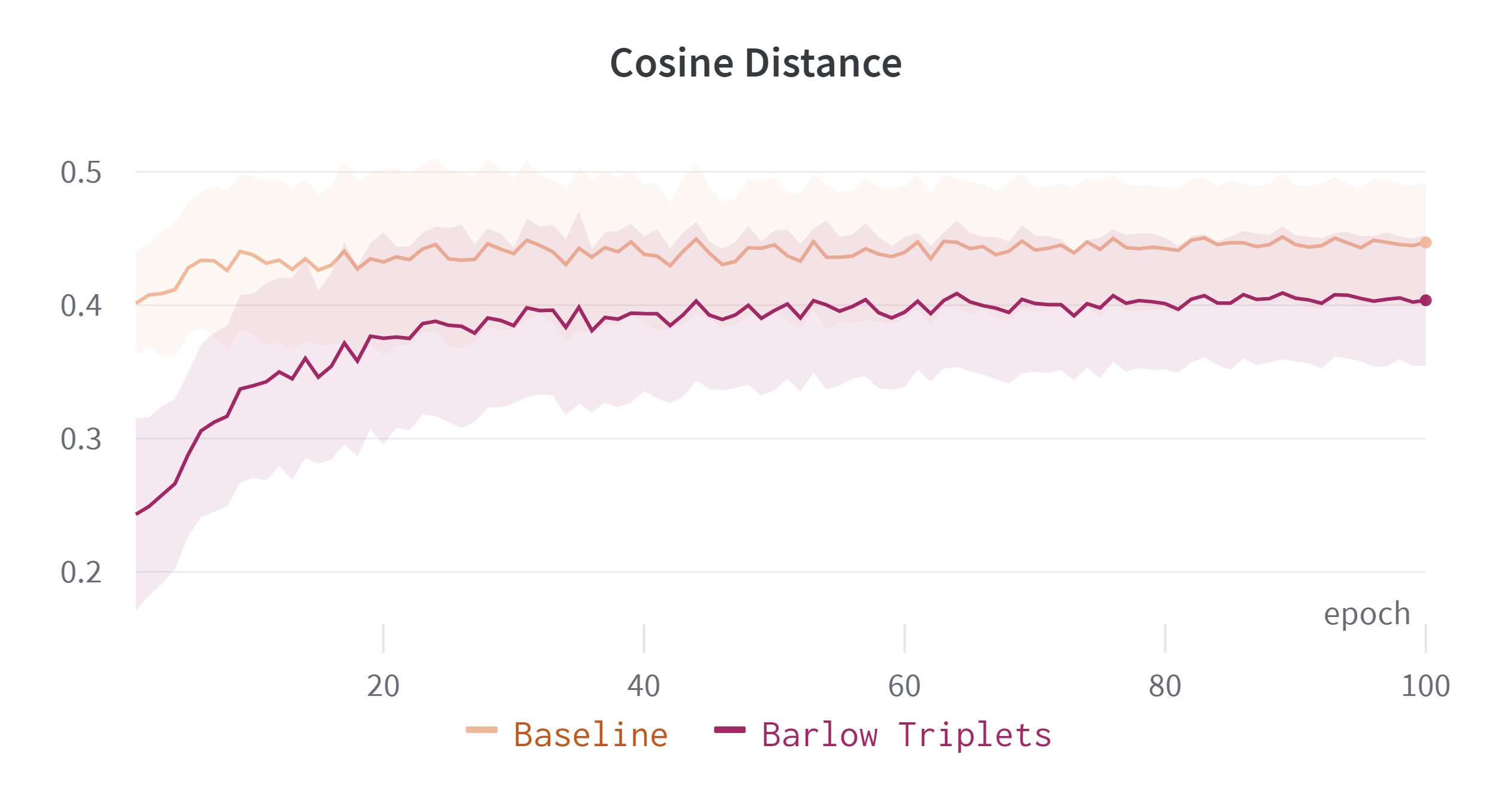}
 \label{subfig:finetuning}
\end{subfigure}
\caption{Cosine distance of CS2 scanner to seen (NZ 210, NZ 2.0) and unseen  (P 1000, GT 450) scanners ($\mu \pm \sigma$ of three repetitions).}
\label{fig:cosine_distance}
\end{figure}

To evaluate whether the \triplets{} successfully aligned the pair-wise scanner representations, we monitored the mean cosine distance of the CS2 embeddings to the other scanners at the encoder bottleneck, as visualized in \cref{subfig:pretraining}. The plot shows a steady decrease especially for the two NanoZoomers incorporated in the self-supervised pre-training, but also for the two scanners that were not seen during training (P 1000 and GT 450). \Cref{subfig:finetuning} compares the mean patch embedding distance to the CS2 patches (averaged across all scanners and across single- and multi-domain experiments) for the second stage of the training pipeline. The plot shows that the pre-trained \triplets{} started with a considerably lower cosine distance than the ImageNet-initialized baseline. Even though during this second stage, the pair-wise cosine distance partially increased again, it converged to an overall lower value than the baseline.

\Cref{tab:performance} summarizes the test set performance of the baseline and the pre-trained model in the single-domain and multi-domain setup of the fully-supervised training stage ($\mu \pm \sigma$ of three repetitions). For inference, we used a moving-window approach with a 128-pixel overlap and center-cropped the segmentation output. We accumulated the confusion matrix across all \acp{wsi} of our test set and calculated the \ac{miou}. The performance of the single-domain baseline highlights the domain shift within the cross-scanner dataset, as the \ac{miou} varies between \num{0.76} and \num{0.84}. The out-of-domain \ac{miou} was increased when using the self-supervised pre-trained \triplets{} in this single-domain setup. Incorporating the two NanoZoomers into the fully-supervised training improved the performance across these two scanners but slightly degraded the performance on the CS2 scanner. Interestingly, the performance on the unseen P1000 scanner was improved the most. The pre-trained \triplets{} improved the baseline performance of the multi-domain setup overall by either increasing the mean performance or reducing the variance.       

\begin{table}[htb]
\begin{adjustbox}
{width=\linewidth}
\begin{tabular}{lS[table-format = 1.2(2)]S[table-format = 1.2(2)]S[table-format = 1.2(2)]S[table-format = 1.2(2)]S[table-format = 1.2(2)]}
\hline
 \textbf{\ac{miou}} & \multicolumn{1}{c}{CS2} & \multicolumn{1}{c}{NZ 210} & \multicolumn{1}{c}{NZ 2.0} & \multicolumn{1}{c}{P 1000} & \multicolumn{1}{c}{GT 450}\\
\hline
Baseline (single) & 0.76(02) & 0.79(02) & 0.84(01) & 0.78(02) & 0.84(02) \\
\triplets{} (single) & 0.75(01) & 0.81(02) & 0.85(01) & 0.79(02) & 0.85(01) \\
Baseline (multi) & 0.71(02) & 0.80(02) & 0.88 (01) & 0.85(01) & 0.83(02) \\
\triplets{} (multi) & 0.75(02) & 0.82(01) & 0.88(00) & 0.85 (01) & 0.83(01) \\
\hline
\end{tabular}
\end{adjustbox}
\caption{\label{tab:results-class} Class-averaged (tumor, non-tumor, background) \acl{miou} ($\mu \pm \sigma$ of three repetitions) of baselines and \triplets{}.}
\label{tab:performance}
\end{table}

\begin{figure}[htb]
\centering
\begin{minipage}[b]{1.0\linewidth}
  \centering
  \centerline{\includegraphics[width=\textwidth]{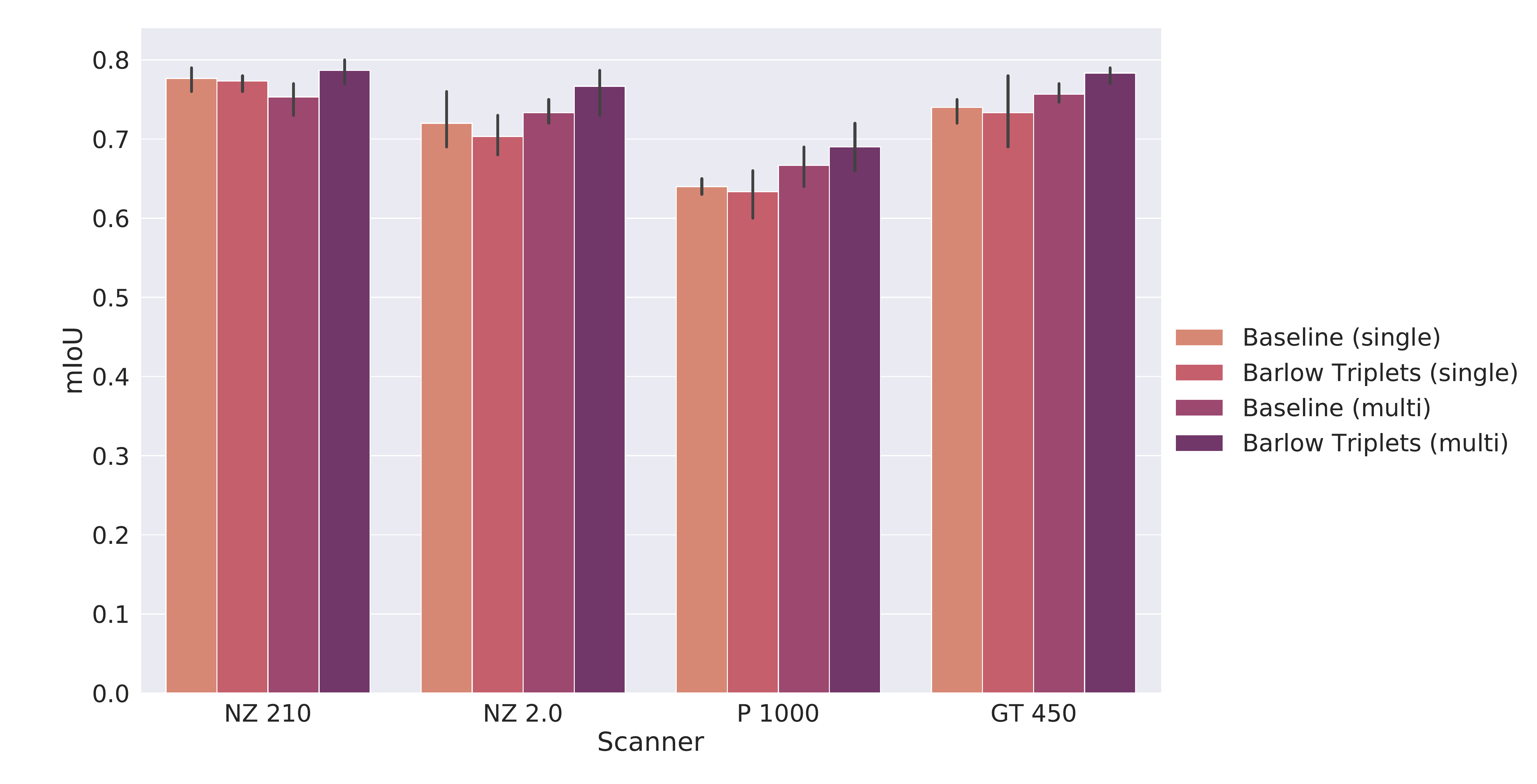}}
\end{minipage}
\caption{Concordance of segmentation outputs to the CS2 prediction measured as \acl{miou} of masks.}
\label{fig:concordance}
\end{figure}

\Cref{fig:concordance} visualizes the agreement of the CS2 prediction and target scanner predictions. For the single-domain setup, the baseline and the pre-trained model yielded a similar averaged concordance across all scanners. The multi-scanner fully-supervised training (Baseline (multi)) helped to align the segmentation outputs of all scanners and self-supervised pre-training increased the concordance even further.     

\section{Discussion}
The large discrepancy in segmentation performance across the different scanners has highlighted the scanner-induced domain shift within our dataset. This gap could neither be completely closed by training the U-Net fully supervised on multiple scanner domains nor by initializing the model with the pre-trained encoder using the \triplets{}.   

During pre-training we observed a continuous decrease of the pair-wise cosine distance at the encoder bottleneck, indicating that the \triplets{} helped to align the scanner representations. When integrating this encoder into a U-Net-like architecture, however, these bottleneck embeddings are passed through the decoder and enriched by features from the skip connection. We here see the potential risk of scanner-specific features being bypassed through the skip connections of the U-Net. Future work could therefore use representation learning strategies on multiple encoder levels.  

Interestingly, all models performed worst on the CS2 scanner, even though this scanner was seen for all experiments (single- and multi-domain). A closer evaluation of the segmentation masks showed that the model considerably underestimated the tumor area on the CS2 \acp{wsi} compared to all other scanners. This indicates that some scanner characteristics were inherently more beneficial for the downstream task of tumor segmentation than others, e.g. different color representations that facilitate the separation of hematoxylin and eosin components or differences in contrast or sharpness. While the results are not fully conclusive, the performance increase in the (multi) setting could indicate that the CS2 indeed benefits from a feature alignment that enhances these characteristics. Overall, we however only observed a slight increase in \ac{miou} compared to the respective baseline. Although the multi-scanner views should fulfill the \textit{InfoMin} principle~\cite{tian2020} and put aside the need for finding effective augmentations~\cite{stacke2021}, the small inter-class variance in histopathology may make it difficult to fully exploit this. Since we observe a stronger effect on the segmentation concordance (\cref{fig:concordance}) than on the performance (\cref{tab:performance}), it will be interesting to understand the effect of label noise on feature alignment.
 
Previous work has mostly employed representation learning as self-supervised pre-training due to limited annotations for downstream tasks. One of the key strengths of the CATCH dataset is its extensive annotation database, which allowed the fully-supervised training of tumor segmentation with high performance on the test set~\cite{wilm2022}. To transfer this performance across different scanner domains, future work could explore a joint or alternating self-supervised training of the encoder and fully-supervised training for the downstream task. 

\section{Conclusion}
In this work, we investigated self-supervised pre-training for scanner-induced domain shifts in histopathology. Our experiments show that self-supervised pre-training is generally applicable to the task of cross-scanner representation alignment but did not yield a significant performance boost for our downstream task. Our results indicate that some scanner-specific characteristics might be relevant for the downstream task of tumor segmentation, which has to be considered when employing representation learning to mitigate scanner-induced domain shifts.   

\vfill
\pagebreak

\section{Compliance with ethical standards}
All specimens were submitted by veterinary clinics or surgeries for routine diagnostic examination of neoplastic disease. As to local regulations, no ethical vote is required for these samples.

\section{Acknowledgments}
F.W. gratefully acknowledges the financial support received by Merck Healthcare KGaA. K.B. gratefully acknowledges support by d.hip campus - Bavarian aim in form of a faculty endowment.

\bibliographystyle{IEEEbib}
\bibliography{strings,refs}

\begin{acronym}
\acro{wsi}[WSI]{whole slide image}
\acro{scc}[SCC]{squamous cell carcinoma}
\acro{he}[H\&E]{Hematoxylin \& Eosin}
\acro{roi}[ROI]{region of interest}
\acro{iou}[IoU]{intersection over union}
\acro{miou}[mIoU]{mean intersection over union}

\end{acronym}

\end{document}